\documentclass[pre,twocolumn,a4paper,showpacs]{revtex4}
\usepackage{amsmath}
\usepackage{amssymb}
\usepackage{amsfonts}
\usepackage{graphicx}
\usepackage{subfigure}
\usepackage{latexsym}
\usepackage{wasysym}
\begin{document}

\title{Solvable model for template coexistence in protocells}

\author{J. F.  Fontanari}
\affiliation{Instituto de F\'{\i}sica de S\~ao Carlos,
  Universidade de S\~ao Paulo,
  Caixa Postal 369, 13560-970 S\~ao Carlos, S\~ao Paulo, Brazil}

\author{M. Serva}
\altaffiliation[On leave of absence from ]{Dipartimento di Ingegneria e Scienze 
dell'Informazione e Matematica, Universit\`a 
dell'Aquila, Italy}

\affiliation{Departamento de Biof\'{\i}sica e Farmacologia,
Universidade Federal do Rio Grande do Norte,
59072-970 Natal, RN, Brazil}


\begin{abstract}
Compartmentalization of self-replicating molecules (templates) in protocells is a necessary step towards the evolution 
of modern cells.   However, coexistence between  distinct template types inside a protocell can  be
achieved only if  there is a selective pressure  favoring protocells with a mixed template
composition. Here we study analytically a group selection model for the coexistence between
two template types using the diffusion approximation of population genetics. The model combines
competition at the  template and  protocell levels as well as genetic drift inside protocells.
At the steady state, we find  a continuous phase transition  separating the
coexistence and segregation regimes, with the order parameter vanishing linearly with the distance to the critical point.
In addition, we derive explicit analytical expressions for the critical  steady-state probability density of protocell compositions.
 \end{abstract}

\maketitle

\section{Introduction}\label{sec:1} 

Explaining the coexistence among selfish individuals is an alluring  issue that pervades  all 
disciplines concerned with the emergence and stability of complex structures evolving
under the rules  of natural selection. A series of classical quandaries
 can be traced back to this issue such as the plankton paradox in Ecology \cite{Hutchinson_61},
the tragedy of commons in Sociology  \cite{Hardin_68} and the information crisis of prebiotic evolution
\cite{Eigen_71},  just to name  a few.  The many ingenious solutions  proposed to tackle those dilemmas, e.g.,
nonequilibrium predator-mediated coexistence \cite{Caswell_78}, coexistence in flows with chaotic mixing \cite{flow_00},
kin selection \cite{Hamilton_64},   reciprocal altruism \cite{Trivers_71}, cyclic cooperative interactions \cite{Eigen_79}  and group selection \cite{Wilson_80}, 
have become major research topics by themselves.

Here we revisit the problem of coexistence of selfish individuals using the group selection framework
in the context of prebiotic evolution \cite{Niesert_81,Szat_87,Fontanari_06,Silvestre_07}. 
We recall that, in this context, coexistence  is hypothesized to circumvent  Eigen's paradox of  the origin of life \cite{Eigen_79} --
 no large genome without enzymes, and no enzymes without a large genome -- by assuming that, initially,  each
short template coded for a small piece of a modular enzyme  (see, e.g., \cite{Manrubia_07}).
The (haploid) individuals are viewed as self-replicating molecules or templates and the groups as  protocells or vesicles that are themselves
capable of reproduction. This is then a prototypical multilevel selection problem \cite{Traulsen_05} which has been analyzed
chiefly through Monte Carlo simulations and numerical iteration of recursion equations. The two dynamics that govern
the competition between  templates and  between protocells  are coupled because the reproduction rate of
the protocells depends on their template composition.

The evolutionary processes we consider  here are individual selection
and group selection. In addition, since we assume that the population of templates inside each protocell is large but finite, random 
genetics drift plays an important role too, especially in  the  hindering of template coexistence. The population of protocells, however, is assumed infinite.
We consider two distinct types of templates only and use the so-called diffusion approximation  
to obtain the partial differential equation that determines the proportion of protocells with a given composition of
templates at a given time \cite{Kimura_83}. Analysis of the (singular) steady-state solution  reveals  
a continuous phase transition separating the coexistence regime, in which  a fraction of protocells exhibit a mixed composition
of templates, from the segregation regime, in which there is no coexistence of template types inside a protocell. More pointedly,
we show that the order parameter vanishes linearly with the distance to the critical line,
and  derive an analytical expression for the  critical steady-state solution.


\section{The model}\label{sec:2}

The population is divided into  an infinite number of protocells each of which carrying exactly $N$
templates. There are two types of templates -- type 1 and type 2 -- which differ only by their replication efficiencies: type 1  has
a selective disadvantage $s > 0$ relative to type 2.  We denote by $x \in \left [ 0, 1 \right ]$ the frequency of type 1 templates 
within a protocell (the frequency of type 2 templates is then $1-x$), and by $\phi \left ( x, t \right )$ the probability density 
of protocells with a fraction $x$ of type 1 templates. Here we assume that $N$ is sufficiently large so that $x$ can be treated as 
a continuous variable. Hence $\phi \left ( x, t \right ) \Delta x $ yields the proportion of protocells carrying 
type 1 templates with frequency  lying in the range $\left ( x, x+\Delta x \right )$.
There is no mutation between the template types and there is no migration 
of templates between protocells. 

Within each protocell, template reproduction follows the rules of the standard Wright-Fisher model
\cite{Crow_70}. In particular,  we assume  that the individual selection coefficients 
 are $1-s$ and  $1$ for type 1 and type 2 templates, respectively. So, if the frequency of  type 1 templates within  a 
given protocell is
$x$ before reproduction, then that frequency  becomes
$\tilde{x}= x \left ( 1-s \right )/\left ( 1-sx \right ) \approx x - sx \left ( 1-x \right )$ after reproduction. 
 We assume, as usual, that $s$ is on the order of $1/N$. This deterministic
process is followed by the random sampling of $N$ templates: the probability that the given protocell
carries exactly    $i=0, \ldots, N$ templates of type 1  is  given by the binomial
\begin{equation}
\binom{N}{i} \tilde{x}^i \left ( 1 - \tilde{x} \right )^{N-i} ,
\end{equation}
so that the frequency of type 1 templates after deterministic selection and random sampling is $x'= i/N$.
The way this procedure changes the probability density $\phi \left ( x, t \right )$ is  derived  using the  diffusion 
approximation of population genetics \cite{Crow_70}, which consists essentially of the calculation of
the first two jump moments $\left \langle \left ( x' - x \right )^n \right \rangle$ to obtain the
drift ($n=1$) and the diffusion ($n=2$) terms of a Fokker-Planck equation (see eq.\ \ref{FP1}).

The competition between protocells is taken into account as follows. Denoting  by $c \left ( x \right )$ the
selection coefficient of a protocell with a fraction $x$ of type 1 templates  we have 
\begin{equation}
\phi \left ( x , t + \Delta t \right ) = \left [ \phi \left ( x , t \right ) + c \left ( x \right )  \phi \left ( x , t \right )  \Delta t \right ] \zeta
\end{equation}
where $\zeta$ is such that $\int_0^1 dx \phi \left ( x , t + \Delta t \right )  =1$, i.e, 
$\zeta = 1/\left ( 1 + \bar{c} \Delta t \right )$ with
\begin{equation}\label{cbar}
\bar{c} = \int_0^1 c \left ( x \right ) \phi \left ( x , t  \right )  dx.
\end{equation}
Finally, taking the limit $\Delta t \to 0$ we obtain the change in the fraction of protocells due to intercell
selection, $\Delta \phi = \left [ c \left ( x \right ) - \bar{c} \right ]  \phi \left ( x , t  \right ) \Delta t$.

Combining the changes in $\phi$ due to the Wright-Fisher process and the intercell selection results
in the equation \cite{Kimura_83}
\begin{equation}\label{FP1}
\frac{\partial}{\partial t}\phi=  \frac{1}{2}\frac{\partial^2}{\partial x^2} \left [ b(x) \phi \right ]
- \frac{\partial}{\partial x} \left [ a(x)\phi \right ]
+ \left [ c(x)-\bar{c} \right ] \phi
\end{equation}
where $b(x) = x \left ( 1 - x \right )/N$,  $a(x) = -s x \left ( 1 - x \right )$, and $\bar{c}$ is given by eq.\ (\ref{cbar}).
 Equation (\ref{FP1}) was first derived by Kimura
aiming at studying the efficiency of group selection on the evolution and maintenance of an altruistic character   \cite{Kimura_83}.
In our notation, the altruists are the type 1 templates, which have a selective disadvantage relative to type 2 or nonaltruistic templates.
In the altruism context, the intercell selection coefficient or group selection pressure must be an increasing function of  $x$, so that the
higher the frequency of altruists in a group, the greater is the  selective advantage of the group. In particular, Kimura has chosen
the linear dependence $c(x) \propto x$  and included the effects of mutation and migration in his analysis \cite{Kimura_83}.
It should be noted, however, that the introduction of mutation and migration, which affects only the drift term $a(x)$ in eq.\ (\ref{FP1}),
actually greatly simplifies the analysis since these mixing processes guarantee the existence of a regular  equilibrium distribution \cite{Wright_31}.

Here we consider the coexistence problem instead, which is considerably more taxing to group selection than the altruistic version
summarized above. In fact,  despite their handicap at the individual  level,  the altruistic  templates have a nonzero probability
of being fixed in small groups solely through the effect of random drift, whereas this very  effect  is a major pressure against the coexistence 
between different templates
\cite{Fontanari_06}.  To favor coexistence we choose the intercell selection coefficient
\begin{equation}\label{presc}
c \left ( x \right ) = c x \left ( 1 - x \right )
\end{equation}
which is maximum for well-balanced protocells at which $x=1/2$. Here $c$ is a parameter on the order of $1/N$
which measures the intensity of the group selection pressure towards coexistence. Prescription (\ref{presc}) is built around the so-called metabolic model of template cooperation which assumes that
the  presence of the two functional template types in the same protocell is needed to assemble a nonspecific replicase which, in turn,
plays an essential role in the template replication process \cite{Niesert_81,Szat_87,Fontanari_06,Silvestre_07}. Since the hookup of the
replicase requires products from the two template types, its production rate is proportional to the concentration of the rare  type, hence
the requirement that $c \left ( x \right )$ is maximized by well-balanced protocells. 
In addition, if a protocell lacks any  template type it is considered unviable,
i.e., $c \left ( 0 \right ) = c \left ( 1 \right ) = 0$.
We note that in the metabolic
model of cooperation each template type contributes indirectly to the replication of the other 
type through the catalytic action of the nonspecific replicase.

\section{Analysis of the steady state}\label{sec:3} 

At the steady state $\partial \phi/\partial t = 0$, eq.\ (\ref{FP1}) reduces to
\begin{eqnarray}\label{FP2}
\frac{d^2}{d x^2} \left[ x \left ( 1-x \right ) \phi \right ]
+S \frac{d}{d x} \left[x \left (1-x \right )\phi \right]&  &\nonumber \\
+ \left [ C x \left (1-x \right )-\bar{C}\right  ]\phi & =  & 0
\end{eqnarray}
where $S = 2Ns$ and $C=2Nc$ are now parameters that can take on arbitrary positive values, and  
\begin{equation}\label{Cbar}
\bar{C}=C\int_0^1 x \left (1-x \right )\phi \left (x \right)d x .
\end{equation}
The solution  $\phi$ has to be found in the interval $[0,1]$ and since the extremes of this
interval  are absorbing barriers it can be written in the general form
\begin{equation}\label{sol2}
\phi \left ( x \right ) = A_0 \, \delta \left( x \right ) + A_1 \, \delta \left (1-x \right ) + B \eta \left( x \right)
\end{equation}
where $A_0, \, A_1$ and $B$ are positive constants (weights) that depend 
on the parameters $S$ and $C$ only. Here
$\eta \left( x \right) $ is a regular function that satisfies the second order differential equation
(\ref{FP2}) in the open interval $\left ( 0, 1 \right )$  and can be eventually continued in the interval 
$ \left [ 0,1 \right ] $ by defining $\eta \left  (0 \right) = \lim_{x \to 0} \; \eta \left  (x \right )$ and
$\eta \left (1 \right ) = \lim_{x \to 1} \; \eta \left (x \right)$. In addition,  imposing the  normalization
$\int_0^1 \eta \left ( x \right ) dx = 1$ we have 
\begin{equation}\label{norm}
A_0 + A_1 +  B = 1 
\end{equation}
 and
\begin{equation}\label{BC}
\bar{C}=BC\int_0^1 x \left (1-x \right )\eta \left (x \right )dx.
\end{equation}
We note that eq.\ (\ref{sol2}) implies that, in the general case,  the  population is  composed of
three types of protocells:  homogeneous protocells carrying  type 1 templates only, homogeneous
protocells carrying type 2 templates only, and heterogeneous protocells carrying any arbitrary
mixture of the two templates. The proportion of each type   in the infinite protocell  population is 
$A_0$,  $A_1$  and  $B$, respectively.

Next  we derive two most useful relations between the values of $\eta \left ( x \right )$ at the extremes
$x=0$ and $x=1$ and the weights that appear in eq.\ (\ref{sol2}). Integrating 
eq.\ (\ref{FP2}) over the interval $\left [-\epsilon, \epsilon \right ]$  and neglecting terms of order of $\epsilon$
and higher 
yield $A_0 = B \eta \left ( 0 \right )/ \bar{C} $ or, equivalently,
\begin{equation}\label{A0}
A_0 = \frac{ \eta \left ( 0 \right ) }{C \int_0^1 x \left (1-x \right ) \eta \left (x \right ) dx},
\end{equation}
where we have used $\frac{d}{dx}
\left[ x(1-x)\eta(x) \right]  = x(1-x) \frac{d}{d x}\eta(x)  
+ (1-2x)\eta(x)$,  
$\eta(\epsilon)= \eta(0) + o(\epsilon)$ and $\eta(-\epsilon)= 0$. Here $o(\epsilon)$ contains all terms of order 
$\epsilon$ or higher. Similarly,  integration of eq.\ (\ref{FP2}) over the interval $\left [1-\epsilon, 1+\epsilon \right ]$ yields 
\begin{equation}\label{A1}
A_1 = \frac{ \eta \left ( 1 \right ) }{C \int_0^1 x \left (1-x \right ) \eta \left (x \right ) dx} .
\end{equation}
We note that  $B=1-A_0-A_1$ can be inserted in eq.\ (\ref{BC}) so that the regular solution $\eta$
can be obtained autonomously (see eq.\ (\ref{BCY}) for a similar operation). 

The numerical procedure to find the regular solution $\eta \left ( x \right )$ of eq.\ (\ref{FP2}) is  greatly facilitated if we define
the  auxiliary function $y \left ( x \right )$ as
\begin{equation}\label{y}
\eta \left ( x \right ) = R \, \exp \left ( -S x /2\right ) \,y \left ( x \right )
\end{equation}
where $R = 1/\int_0^1  dx \exp \left (- Sx/2\right )  y \left  ( x \right )  $ 
 is introduced to  guarantee the normalization of $\eta$, leaving us
 free to  impose the normalization  $\int_0^1  dx  \,  y (x) = 1$ on the auxiliary function  $y$.
It is easy to verify from eq.\ (\ref{FP2}) that in the open interval  $\left (0,1 \right )$ we have
\begin{equation}\label{gq}
\frac{d^2}{d x^2}
\left[ x \left ( 1-x \right ) y \right]
+ \Gamma x \left (1-x \right ) y = \bar{C} y  
\end{equation}
where $\Gamma=C- S^2/4$ and $\bar{C}$ is obtained by integration of
eq.\  (\ref{gq}) in the interval $\left ( 0, 1 \right )$ as
\begin{equation}\label{BCY}
\bar{C} \, = \, \Gamma   \int_0^1  x \left ( 1- x \right ) y \left ( x \right ) dx 
 -  \,y \left ( 0 \right )  -  \,y \left ( 1 \right ) 
\end{equation}
where $y(0) =  \lim_{x \to 0} y(x)$  and  $y(1) = \lim_{x \to 1} y(x)$. Since the differential equation  (\ref{gq}) is symmetrical around 
$x=1/2$, we assume that the solution $y \left (x \right )$ is symmetrical too, so  that $y(0) = y(1)$. 

\section{Numerical Solution}\label{sec:4} 

At this stage our mathematical problem is ready for a numerical approach. 
We are left  with a Sturm-Liouville problem, eq.\ (\ref{gq}),  without boundary conditions  which can be solved by requiring the regularity of $y \left ( x \right )$ in $\left [0, 1 \right ]$ only \cite{Chalub_09}. Of course, this requirement is satisfied provided that $\bar{C}$ is the eigenvalue of the Sturm-Liouville problem, which   depends solely on the  value of $\Gamma$.
In practice we determine $\bar{C}$ by propagating the solution using the Runge-Kutta algorithm from  $x=0$ to $x=1/2$, from $x=1$ to $x=1/2$, 
and then requiring that the two solutions join smoothly at $x=1/2$ (see page 746 of \cite{NR}). 
We use an arbitrary value for $y(0)=y(1)$ to begin the Runge-Kutta
iterations, and the first  derivatives $y'(0) = -y'(1) = \left ( 1 + \bar{C}/2 \right ) y (0)$  as given by eq.\ (\ref{gq}). 
The arbitrariness of the initial condition  has no effect on the resulting eigenvalue $\bar{C}$ and it is mitigated by imposing the normalization condition on $y(x)$.
In fact,   since
the coefficients $A_0$ and  $A_1$  given by  eqs. (\ref{A0}) and (\ref{A1}), as well as $B = 1 - A_0 - A_1$,  depend  on the ratio of terms 
involving $\eta \left ( x \right )$ (and so  $y \left ( x \right )$)  our arbitrary choice of $y(0)$
 is inconsequential.  Finally, we note that eq.\ (\ref{BCY})  does not provide any additional information about
$y$ or $\bar{C}$ -- it is derived from eq.\ (\ref{gq}) -- and so it is not used in our numerical procedure.

\begin{figure}[!t]
 \begin{center}
\subfigure{\includegraphics[width=1.0\linewidth]{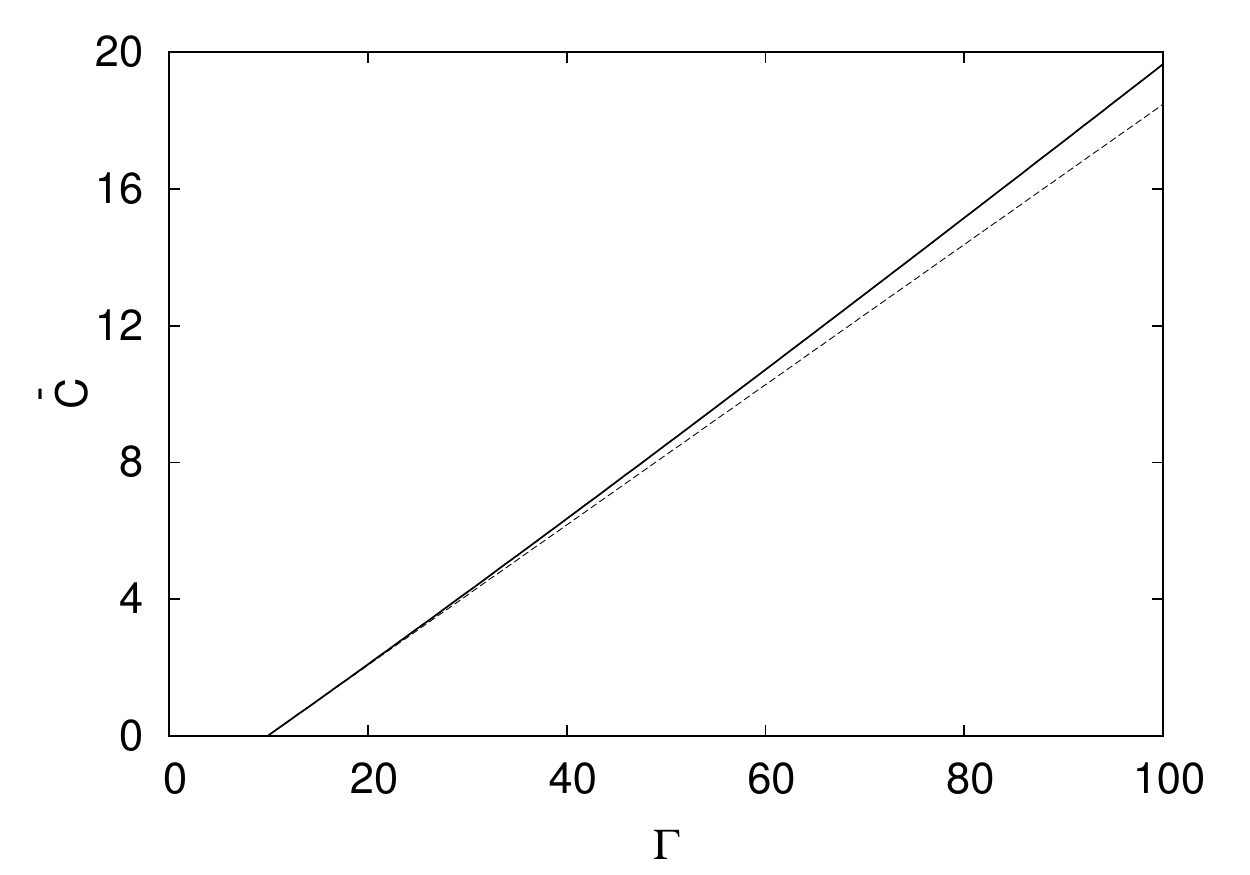}}
\subfigure{\includegraphics[width=1.0\linewidth]{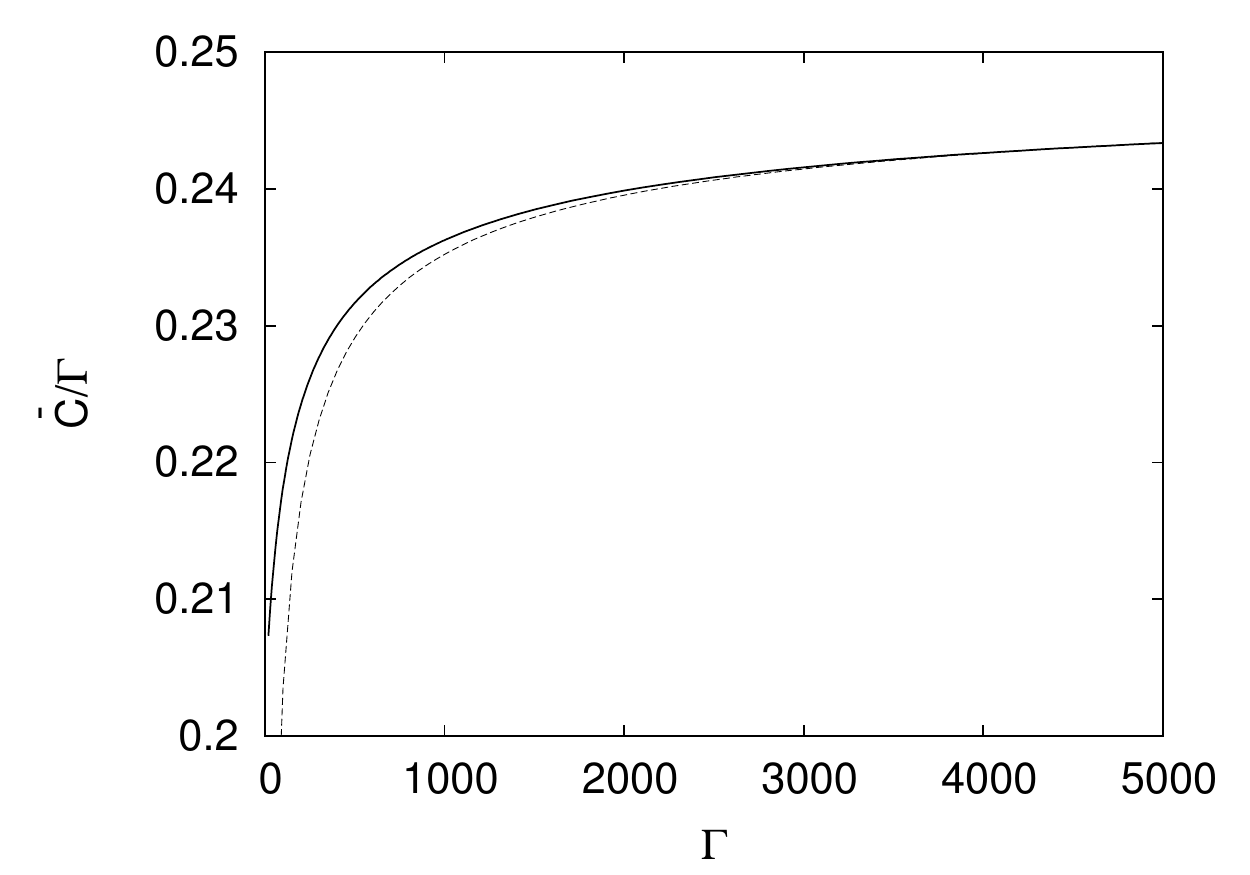}}
  \end{center}
\caption{Eigenvalue $\bar{C}$  as function of $\Gamma = C - S^2/4$. For
$\Gamma \leq \Gamma_c = \pi^2$ we find $\bar{C} =0$,  which characterizes 
the segregation regime. The dashed line in the upper panel is
$\bar{C} = \alpha \left ( \Gamma - \Gamma_c \right )$ with $\alpha$ given by eq.\ (\ref{alpha}).
The lower panel shows the ratio $\bar{C}/\Gamma$ in a magnified scale of $\Gamma$, where the dashed
line is the fitting $\bar{C}/\Gamma = 0.25 - 0.47/\Gamma^{1/2}$.}
\label{fig:1}
\end{figure}

The dependence of the  eigenvalue $\bar{C}$ on  $\Gamma$ exhibited in the upper panel of fig.\ \ref{fig:1} reveals the existence of
a critical value $\Gamma_c$  below which $\bar{C}=0$. This critical point separates two distinct
steady-state  regimes  regarding  the possibility of coexistence of the 
two types of templates within the same  protocell, and so $\bar{C}$ can be seen as the order parameter
of our problem. We note that  $\bar{C}/\Gamma$ (the coefficient of the linear term
in the expansion of  $\bar{C}$ in powers of $\Gamma$) varies very slowly with increasing  $\Gamma$ (see lower panel of fig.\ \ref{fig:1}),
being confined to the interval $\left [ 0.205, 0.25 \right ]$. In fact, since in the limit of large $\Gamma$ we have
$\Gamma \approx C \gg 1$ the group selection pressure favoring coexistence becomes the dominant force
leading to an ideal coexistence scenario, i.e.,  $\phi = \delta \left ( x - 1/2 \right )$, for which
 $\bar{C}/\Gamma \approx  1/4$ (see eq.\ (\ref{Cbar})). 
 
Before we carry on the characterization of the two  steady-state regimes separated by $\Gamma_c$, let us
show how  the value of this critical  parameter can be calculated.
 
 \section{The transition line}\label{sec:5}

Setting $\bar{C}=0$ and defining $\psi \left (x \right ) = x \left ( 1-x \right ) y \left (x \right ) $, 
eq.\ (\ref{gq}) reduces to the harmonic oscillator equation   $d^2 \psi/d x^2   + \Gamma_c \psi = 0$.
The solution $\psi =   \kappa_c  \cos \left ( \Gamma_c^{1/2} x \right ) +  \kappa_s \sin \left (  \Gamma_c^{1/2} x \right )$ must vanish
at $x=0$ and $x=1$, otherwise $y \left (x \right ) $ would not be normalizable. This implies that $\kappa_c=0$,
and
$\Gamma_c = \pi^2$ or, equivalently,
\begin{equation}\label{C_c}  
C_c = \pi^2 + \frac{S^2}{4}.
\end{equation}
Hence, at the critical point we have $y_c \left ( x \right )  = \kappa_s \sin \left ( \pi x \right )/\left [ x \left ( 1-x \right ) \right ] $
where the normalization factor is
$\kappa_s = 1/ \int_0^1 dx \sin \left ( \pi x \right )/\left [ x \left ( 1-x \right ) \right ]  \approx 0.270 $ . Finally, returning to the original
regular probability density we write
\begin{equation}\label{Eta_c}
\eta_c \left ( x \right ) = R_c \kappa_s \exp \left ( - Sx/2 \right ) \frac{\sin \left ( \pi x \right )}{x \left ( 1 - x \right )}
\end{equation}
where $R_c$ is  the normalization constant that appear in eq.\ (\ref{y}) evaluated at $\Gamma = \Gamma_c$. 

As a consistency check we will show  now that $\bar{C} = 0$ at $C=C_c$. In fact, eq.\ (\ref{Eta_c}) yields 
$\eta_c \left ( 0 \right ) = R_c \kappa_s  \pi$, $\eta_c \left ( 1 \right ) = R_c  \kappa_s  \pi \exp \left ( - S/2 \right )$, and
\begin{equation}\label{integ}
\int_0^1 x \left ( 1 - x \right ) \eta_c \left ( x \right ) dx = \frac{R_c  \kappa_s  \pi}{C_c} \left [1 + \exp \left ( - S/2 \right ) \right ] .
\end{equation}
Inserting these results in eqs.\ (\ref{A0}) and (\ref{A1}) yields
\begin{equation}\label{A0c}
A_0^c = \frac{1}{ 1 + \exp \left (  -S/2 \right )} 
\end{equation}
and
\begin{equation}\label{A1c}
A_1^c = \frac{\exp \left (  -S/2 \right )}{ 1 + \exp \left (  -S/2 \right )} .
\end{equation}
Since $A_0^c + A_1^c =1$ it follows from eqs.\ (\ref{norm}) and (\ref{BC}) that $B^c=0$  and $\bar{C}=0$.

\section{Characterization of the steady-state regimes}\label{sec:6}

The first regime is associated to the parameter region   $C \leq C_c$,  and 
is  characterized by $\bar{C} = 0$.  According to eq.\ (\ref{Cbar}), its protocell distribution $\phi$ is a sum of Dirac deltas centered at $x=0$ and $x=1$, i.e., $B=0$ in eq.\ (\ref{sol2}). 
Since the two types of templates do not coexist within the same protocell  we refer to this regime as the segregation regime.
This is a non-ergodic regime  where the
weights $A_1$ and $A_0 = 1 - A_1$ depend on the initial conditions, i.e., on the distribution $\phi \left (x,0 \right )$. In the
following we calculate this dependence explicitly in the case $S=0$.

Conforming to the previous rescalings $C=2Nc$ and $S=2Ns$, we begin by replacing $ t $  by  $2Nt $  in eq.\ (\ref{FP1}). In addition,
to lighten the notation  we introduce the  abbreviation
 $\left \langle f  \left ( x \right ) \right \rangle_t = \int_0^1 f \left ( x \right ) \phi \left ( x ,t\right ) dx$ for 
the expected value of  a regular function  $f \left ( x \right )$ at time $t$. Hence
\begin{eqnarray}\label{s_f1}
\frac{d}{dt} \left \langle f  \left ( x \right ) \right \rangle_t  & = & \left \langle x \left ( 1- x \right ) \frac{\partial^2 f (x)}{\partial x^2} \right \rangle_t
- \bar{C} \left ( t \right ) \left \langle f  \left ( x \right ) \right \rangle_t
  \nonumber \\
&  &  + C \left \langle x \left ( 1- x \right ) f  \left ( x \right ) \right \rangle_t
\end{eqnarray}
with  $\bar{C} \left ( t \right )  = C \left \langle x \left ( 1- x \right ) \right \rangle_t$. We recall that
$\lim_{t \to \infty} \bar{C} \left ( t \right ) = \bar{C} = 0$ in the segregation regime. The choice 
$  f  \left ( x \right ) = \sin \left ( C^{1/2} x + \theta \right ) $ where $\theta$ is an arbitrary constant allows us to rewrite
eq.\ (\ref{s_f1}) as
\begin{equation}\label{s_f2}
\frac{d}{dt} \left \langle \sin \left ( C^{1/2} x + \theta \right )   \right \rangle_t  = 
- \bar{C} \left ( t \right ) \left \langle \sin \left ( C^{1/2} x + \theta \right ) \right \rangle_t
\end{equation}
whose formal solution is
\begin{equation}\label{s_f2}
\frac{\left \langle \sin \left ( C^{1/2} x + \theta \right ) \right \rangle_t }{ \left \langle \sin \left ( C^{1/2} x + \theta \right ) \right \rangle_0 } =
\exp \left [ - \int_0^t \bar{C} \left ( \tau \right ) d \tau \right ] .
\end{equation}
Since the right-hand side of this equation does not depend on $\theta$, neither does the  left-hand side so we can write
\begin{equation}\label{s_f3}
\frac{\left \langle \sin \left ( C^{1/2} x \right ) \right \rangle_t }{ \left \langle \sin \left ( C^{1/2} x \right ) \right \rangle_0 } 
= \frac{\left \langle \cos \left [ C^{1/2} \left ( x - 1/2 \right ) \right ] \right \rangle_t }{ \left \langle \cos  \left [ C^{1/2} \left ( x - 1/2 \right ) \right ]  \right \rangle_0 }
\end{equation}
for the choices $\theta = 0$ and $\theta = \left ( \pi - C^{1/2}\right )/2$. This identity holds for any $t$ and therefore it holds also
in the limit $t \to \infty$ at which $\lim_{t \to \infty} \phi \left ( x, t \right ) = \phi \left ( x \right ) = A_0 \delta \left ( x\right )
+ A_1 \delta \left ( x - 1 \right )  $.  Evaluation of  the averages in this limit  and use of  $A_0 = 1- A_1$ yield
\begin{equation}\label{s_f4}
A_1 = \frac{ \cos \left ( C^{1/2} /2 \right ) \int_0^1 \sin \left ( C^{1/2} x \right ) \phi \left ( x,0 \right ) dx} 
{\sin \left ( C^{1/2}\right )  \int_0^1 \cos  \left [ C^{1/2} \left ( x - 1/2 \right ) \right ]  \phi \left ( x,0 \right ) dx  }
\end{equation}
where we have reverted to the original integral notation to emphasize the dependence on $\phi \left ( x,0 \right )$.
The classic result $A_1 =\int_0^1  x  \phi \left ( x,0 \right ) dx  $ is recovered for $C \to 0$ \cite{Crow_70} as well as $A_1 = A_1^c = 1/2$ for
$C \to \pi^2$ (see eq.\ (\ref{A1c})).  Hence $A_1$ is continuous at the critical point $C=C_c$. 
Although this approach allows the calculation of the weight $A_1$  for $S>0$ as well, the derivation is too lengthy and will be omitted here.

\begin{figure}[!t]
 \begin{center}
\includegraphics[width=1.0\linewidth]{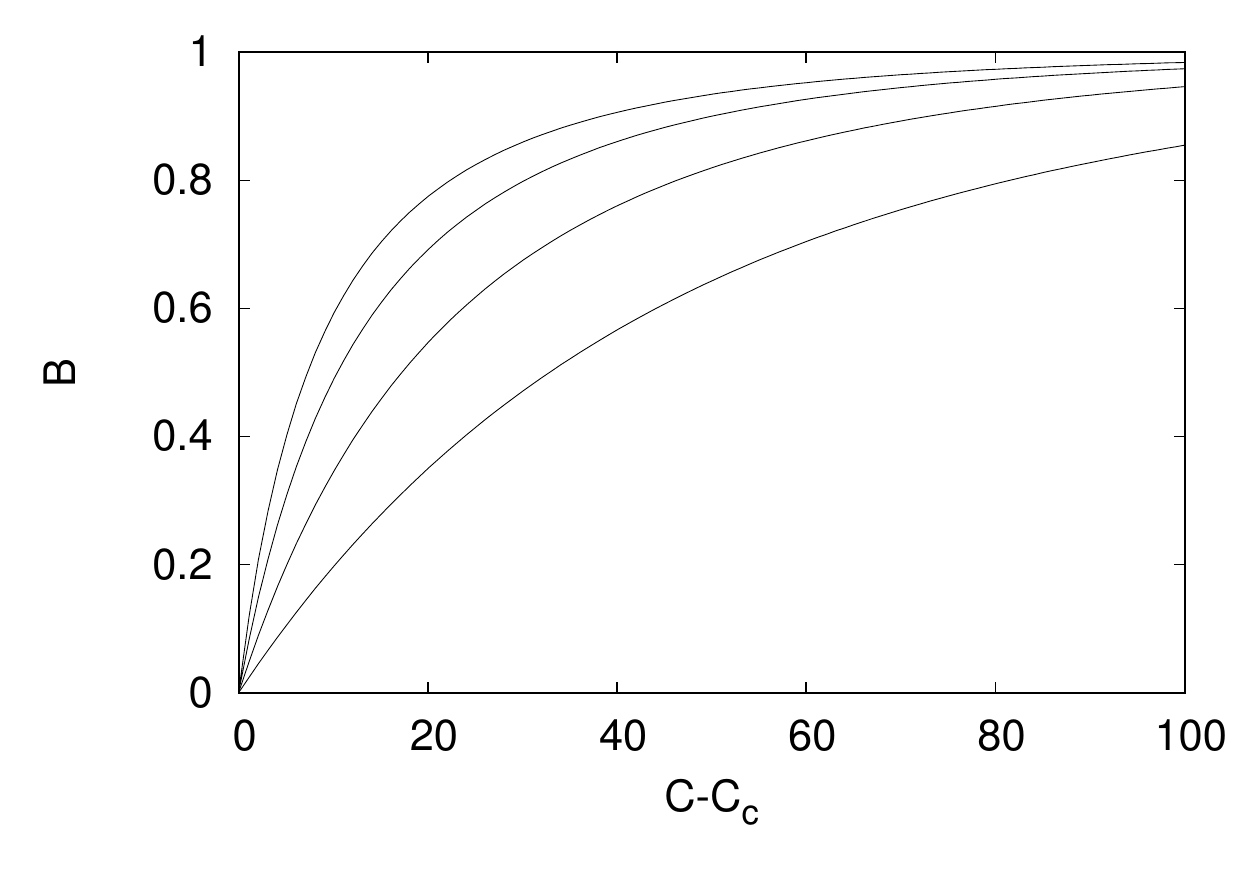}
  \end{center}
\caption{Fraction of protocells carrying the  two template types as function of the
distance to the critical point $C - C_c = \Gamma - \Gamma_c $ for (top to bottom) $S=0, 5, 10$ and
$20$.}
\label{fig:2}
\end{figure}

The second steady-state regime is  characterized by $\bar{C} > 0$ and so admits the existence of protocells in which the two types 
of templates coexist. The quantity of interest here is the weight  $B$ 
(see eq.\  (\ref{sol2})) which gives the fraction of protocells carrying the two  template types, regardless of their frequencies
inside the protocells. In 
fig.\ \ref{fig:2} we present  the dependence of $B$ on the distance to the critical point $\Gamma - \Gamma_c = C - C_c$
for different values of the (rescaled)  individual selection coefficient $S$. Note that although $\bar{C}$ depends on $\Gamma$
only, the weights $A_0$, $A_1$ and $B$ associated to the different protocell types exhibit an  explicit dependence on $S$ as well,
which comes through the function $\eta$  (see eq.\ (\ref{y})). In particular,  for a fixed group selection pressure, increase of
the selective advantage of type 2 templates reduces the fraction $B$ of protocells that exhibit coexistence between templates, as 
expected. More pointedly, we will show  that, close to the critical point, $B$  decreases with $1/S$ for increasing $S$.

\begin{figure}[!t]
 \begin{center}
\includegraphics[width=1.0\linewidth]{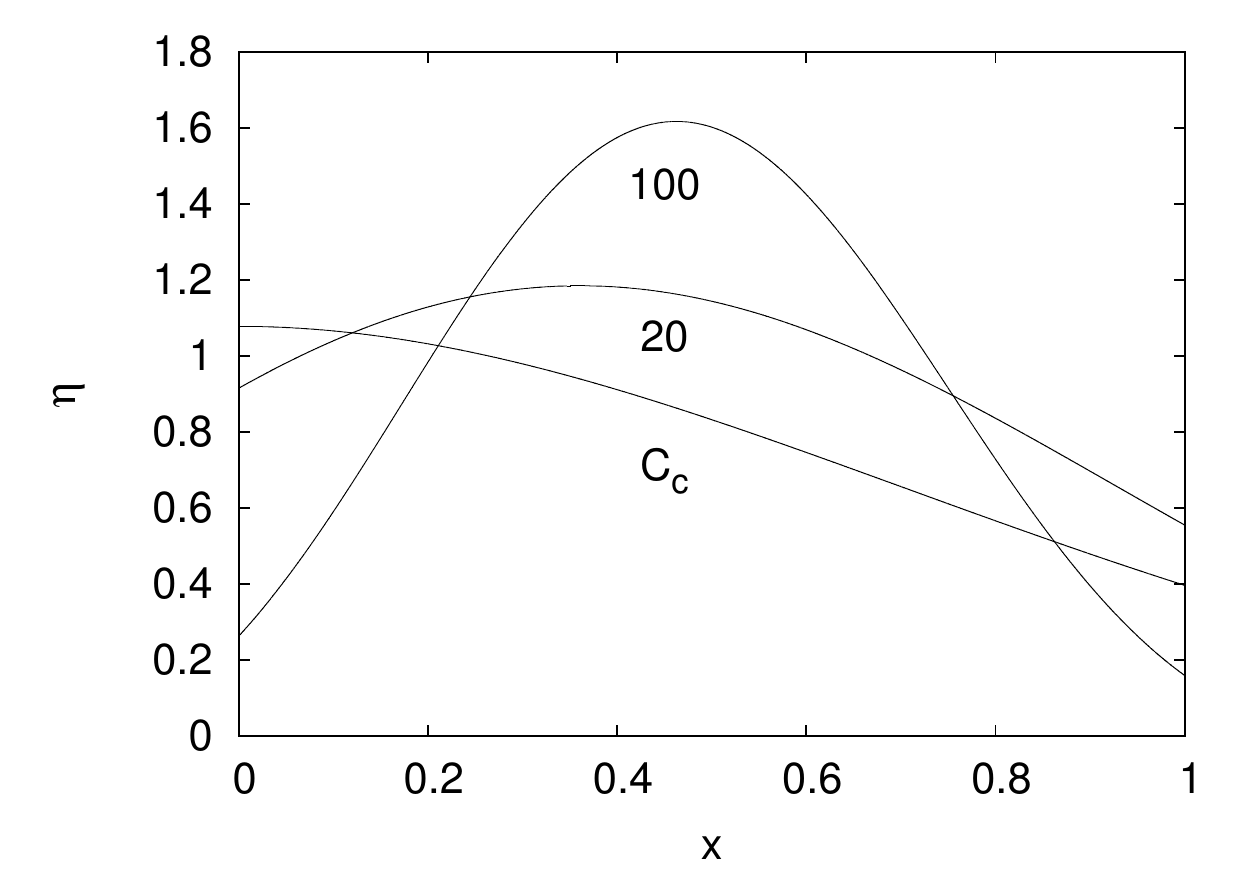}
  \end{center}
\caption{Regular normalized solution $\eta \left ( x \right )$ for $S=1$ and
$C=C_c \approx 10.12$, $20$ and $100$ as indicated in the figure.}
\label{fig:3}
\end{figure}

In fig.\ \ref{fig:3} we show the regular part of the density of probability of protocells  $\eta \left ( x \right ) $ for different group selection intensities. 
The protocells are unbalanced in this figure because of the  choice  $S=1$ which confers a selective advantage to type 2 templates in the 
 competition at the individual level that goes on inside the protocells. However, as $C$ increases
this unbalance diminishes  and eventually the population is completely dominated by well-balanced protocells so 
that $\eta \left ( x \right )  \to \delta \left ( x - 1/2 \right )$.

Next we derive  explicit analytical expressions for the behavior of the order parameter $\bar{C}$ as
well as for the coexistence probability $B$ near the critical point $\Gamma_c$.  
Multiplying eq.\  (\ref{gq}) by $\exp \left ( -\lambda x \right )$ and integrating
in the interval $\left ( 0, 1 \right )$ yield
\begin{eqnarray}
\bar{C}/Q  &  = &  \left ( \Gamma + \lambda^2 \right )   \int_0^1 \exp \left ( -\lambda x \right ) x \left ( 1- x \right ) y \left ( x \right ) dx \nonumber \\
&   &  - y \left ( 0 \right ) \left [ 1 + \exp \left ( - \lambda \right ) \right ] 
\end{eqnarray}
with $Q = 1/\int_0^1 dx  \exp \left (- \lambda x \right )  y \left  ( x \right ) $.
This equation reduces to  eq.\ (\ref{BCY}) when $\lambda=0$ whereas for $\lambda=S/2$  
it coincides with the relation  that is obtained by combining  eqs.\ (\ref{norm}), (\ref{BC}), (\ref{A0}), (\ref{A1}) and (\ref{y})
and taking into account that $\Gamma=C- S^2/4$.

Taking $\lambda = i \pi$ and recalling that $\Gamma_c = \pi^2$ we obtain
\begin{equation}
\bar{C} = \left ( \Gamma - \Gamma_c \right ) \frac{ \int_0^1 x \left ( 1- x \right ) \sin \left ( \pi x \right ) y \left ( x \right ) dx }{\int_0^1  \sin \left ( \pi x \right ) y \left ( x \right ) dx }
\end{equation}
where we have  used  the  fact that $y \left ( x \right )$ is symmetric about $x=1/2$. For $\Gamma$ close to $\Gamma_c$ we can replace $y$ by
$y_c $  and write
$\bar{C} \approx \alpha \left ( \Gamma - \Gamma_c \right )$ where
\begin{equation}\label{alpha}
\alpha =\left [  2 \int_0^1 \frac{\sin^2 \left ( \pi x  \right )}{x \left ( 1- x \right )} dx \right ]^{-1} \approx 0.205 .
\end{equation}
Now it is easy to obtain the behavior of the coexistence  probability $B$ near the critical point. In  fact, once
the behavior of $\bar{C}$ is known, use of eqs.\ (\ref{BC}) and (\ref{integ}) allows us to write
 $B \approx \beta \left ( \Gamma - \Gamma_c \right )$ where
\begin{equation}
\beta = \frac{ \alpha }{\pi \left [ 1 + \exp \left ( - S/2 \right ) \right ]} \int_0^1  \exp \left ( - Sx/2 \right ) \frac{\sin \left ( \pi x \right )}
{x \left ( 1- x \right )}  dx.
\end{equation}
For $S=0$ we have $\beta = \alpha/2 \pi \kappa_s \approx 0.121$ whereas for $S \gg 1$  we find $\beta \approx 2 \alpha/S$.

\section{Conclusion}\label{sec:7}

Group selection and, more generally, structured population arguments aiming at  explaining altruistic
behavior and eusociality in nature have been a source of controversy  since they were first proposed
in the 1960s \cite{Vero_62,Williams_66} (see \cite{Nowak_10}  and accompanying refutations for a recent clash).
However, use of group selection ideas to explain the coexistence of selfish individuals is a much less loaded issue, at least within the  
prebiotic evolution context, since there seems to exist  a consensus that the compartmentalization of  templates was a necessary 
evolutionary step  towards the modern cell \cite{Bresch_80,Eigen_80}.
Regardless of the relevance and controversies surrounding
the role of group selection  in nature, the challenging mathematical models used to describe the resulting 
multilevel selection problem have been viewed as an attraction on their own \cite{Kimura_83,Eshel_72,Aoki_82,Donato_97}. 
Here we follow this tradition and offer a numerical and  analytical solution to a nontrivial group selection model.

We build on the work of Kimura \cite{Kimura_83} and consider a group selection pressure towards
the coexistence of two types of templates with distinct replication rates. We find an explicit
expression for the critical  (i.e., minimum) intensity of group selection   needed to balance the
segregation effects of genetic drift and individual selection, thus guaranteeing 
the existence of protocells carrying the two template types (see eq.\  (\ref{C_c})). In addition, we derive analytical 
expressions for the steady-state distribution at the critical point and show that the transition
between the regimes of coexistence and segregation is continuous with the order
parameter vanishing linearly with the distance to the critical point. The model is nontrivial because it  preserves the
main effect of finite populations -- genetic drift --  which is revealed when the values
of the rescaled parameters $C=2N c$ and $S=2N s$   decrease towards zero.  Since $C_c \to \pi^2$ in this case, template coexistence
is ruled out for small protocells due to the segregating effect of genetic drift.

We note that the actual value of the population size $N$ has little relevance in the diffusion approximation framework,
provided it is large enough to allow treating the frequency $x$  as a continuous variable. The 
relevant quantities that appear in the equations are always expressed
as products between $N$ and the  original parameters ($s$ and $c$ in our case), 
and  only them are accessible to experiments involving small population samples (see Chapter 13 of \cite{Nei_87}).
Nevertheless, we can speculate on the value of $N$  using the
recent estimate that a human cell infected with the  retrovirus HIV-1  may produce more than $10^4$ new viral 
particles over its short life span \cite{HIV}.  This figure may  then be viewed as a rough estimate
of the carrying capacity of the protocells. 

From a mathematical perspective, our work departs from previous population genetic studies 
using the diffusion approximation due to the absence of the mixing and regularizing 
processes of mutation and migration \cite{Crow_70}. Inclusion of these processes would 
guarantee the existence of a regular equilibrium distribution for the Fokker-Planck-like equation (\ref{FP1}).
Use of  prescription  (\ref{sol2}), however, allows us to single out the singularities at the absorbing
barriers $x=0$ and
$x=1$, whereas the relations (\ref{A0}) and (\ref{A1}) provide a link  between the regular part of the
equilibrium distribution and the weights of the  singular parts. This solution strategy can be applied  to
a variety of problems characterized by  the superposition of absorbing and extended steady states.

\acknowledgments
The authors are grateful to Luca Peliti for having spurred their collaboration 
and for having highlighted the issues yet to be clarified about the 
role of altruism in evolutionary processes.
The research of J.F.F. was supported in part by Conselho Nacional de Desenvolvimento
Cient\'{\i}fico e Tecnol\'ogico (CNPq) while the research of M.S. was 
partially supported by PRIN 2009 protocollo n.2009TA2595.02.

\end{document}